\documentclass[draft]{agujournal2019}
\usepackage{url} 
\usepackage{lineno}
\usepackage{subfigure}
\usepackage[inline]{trackchanges} 
\usepackage{soul}
\usepackage{mathtools}
\usepackage{bm}
\usepackage{amsmath,amssymb,latexsym}  
\usepackage{textcomp}
\usepackage[english]{babel}

\renewcommand{\textwidth}{15cm}

\journalname{JGR: Planets}

\begin{document}

%
%

\title{Viscous dissipation in the fluid core of the Moon}

%
%




\authors{Jiarui Zhang \affil{1} and Mathieu Dumberry \affil{1}}

\affiliation{1}{Department of Physics, University of Alberta, Edmonton, Alberta, Canada.}




\correspondingauthor{Mathieu Dumberry}{dumberry@ualberta.ca}



 \begin{keypoints}
 \item The misaligned spin axes of the lunar mantle, fluid core and inner core induce viscous friction at the boundaries of the fluid core. 
 \item For an inner core radius $>80$ km and a free inner core nutation period close to 18.6 yr, friction at the inner core boundary dominates.
\item Viscous dissipation at the ICB is likely insufficient to have ever been above the threshold to power a thermally driven dynamo.
 \end{keypoints}

%
%

%
%

\begin{abstract}
The spin axes of the mantle, fluid core and solid inner core of the Moon precess at frequency $\Omega_p=2\pi/18.6$ yr$^{-1}$ though with different orientations, leading to viscous friction at the core-mantle boundary (CMB) and inner core boundary (ICB).  Here, we use a rotational model of the Moon with a range of inner core and outer core radii to investigate the relative importance of viscous dissipation at the CMB and ICB, and to show how this dissipation is connected to the phase lead angle ($\phi_p$) of the mantle ahead of its Cassini state.  We show that when the inner core radius is $>80$ km and the free inner core nutation frequency $\Omega_{ficn}$ approaches $\Omega_p$, viscous dissipation at the ICB can be comparable to that at the CMB, and in the most extreme cases exceed it by as much as a factor 10.  If so, the viscous dissipation in the lunar core projected back in time depends on how $\Omega_{ficn}$ has evolved relative to $\Omega_p$. We further show that constraints on the CMB and ICB radii of the lunar core can in principle be extracted by matching the observed phase lead of $\phi_p=0.27$ arcsec; this requires an improved estimate of tidal dissipation and an accurate model of the turbulent viscous torque.  Lastly, when our rotational model is constrained to match $\phi_p=0.27$ arcsec, our results suggest that the viscous dissipation at the ICB is likely insufficient to have ever been above the threshold to power a thermally driven dynamo.
\end{abstract}

\noindent {\bf Plain language summary}:
Just like a spinning top, the spin axis of the Moon is precessing in space at period of 18.6 yr.  The spin axes of its fluid core and, if present, its solid inner core precess at the same rate but with different orientations. Here, we calculate the viscous friction at the core-mantle boundary (CMB) and inner core boundary (ICB) induced by this differential rotation for a range of inner core and outer core radii. We show that when the inner core is $> 80$ km, viscous friction at the ICB can be large while that at the CMB is significantly reduced for some lunar models.  Although the exact radii of the solid inner core and fluid core of the Moon are not known, we show how additional information about the core geometry can in principle be extracted by ensuring that the total dissipation is consistent with the observed orientation of the lunar rotation axis in space.  Lastly, our results suggest that convective flows in the liquid core that may be driven by the heat released by viscous friction at the ICB are likely not sufficiently vigorous to generate a magnetic field today, or at any point in the lunar past. 

\section{Introduction}

Tracking of the position and orientation of the Moon by Lunar Laser Ranging (LLR) has revealed a wealth of knowledge on its orbit and rotation \cite<e.g.>{dickey94}, thereby providing important constraints on its interior structure \cite<e.g.>{williams14}.  The orbit normal is inclined by an angle $I=5.145^\circ$ with respect to the ecliptic normal. The spin-symmetry axis is tilted by an angle $\theta_p=1.543^\circ$, also with respect to the ecliptic normal, in the same plane as that formed by the orbit and ecliptic normals, but in the reverse direction, such that the lunar obliquity is $I+\theta_p = 6.688^\circ$.  The orbit normal and spin-symmetry axis are both precessing about the ecliptic normal, in the retrograde sense, with a common frequency $\Omega_p=2\pi/18.6$ yr$^{-1}$, such that they remain coplanar.  This configuration describes a Cassini state \cite{colombo66,peale69}. 

It is convenient to refer to the plane that contains the orbit and ecliptic normals as the Cassini plane.   LLR observations have shown that the spin-symmetry axis does not lie exactly in the Cassini plane, but leads ahead of it by a small angle of $\phi_p=0.27$ arcsec.  This phase lead is indicative of rotational energy dissipation.  Sources of dissipation include viscoelastic tidal deformation \cite{yoder79,cappallo81}, viscous relaxation within a possible solid inner core \cite<>[henceforth referred-to as OD21]{organowski20} and viscous friction at the core-mantle boundary (CMB) \cite{yoder81,williams01}.  Here, we focus on the latter.

The tilt angle $\theta_p=1.543^\circ$ characterizes the orientation of the spin-symmetry vector of the solid outer shell of the Moon comprised of its mantle and crust.  The rotation vector of the fluid core is also precessing at frequency $\Omega_p$, although its tilt angle is different than that of the mantle because the ellipticity of the lunar CMB is too small to provide an inertial coupling sufficiently strong to bring them into alignment \cite{goldreich67}.  No direct observation on the orientation of the spin vector of the fluid core is available, but it should remain close to, though not exactly aligned with, the ecliptic normal \cite<>[the latter two studies are henceforth referred to as DW16 and SD18, respectively]{williams01,meyer11,dumberry16,stys18}.  The differential rotation of the mantle and fluid core leads to viscous friction at the CMB, dissipating rotational energy.   

A fit between LLR observations and a model of lunar deformation and rotation allows to separate the relative contributions of the total dissipation from tidal deformation and CMB friction \cite{williams01,williams15}.  Tidal deformation contributes approximately 0.15 arcsec to the observed 0.27 arcsec phase lead (e.g. OD20).  The rotational model used in LLR studies does not include an inner core, so the remaining 0.12 arcsec is entirely absorbed by viscous friction at the CMB.  If a solid inner core is present, viscous friction also takes place at the inner core boundary (ICB) and contributes to the non-tidal part of the dissipation. 

The estimate of the present-day dissipation at the CMB, $Q_{cmb}$, retrieved from LLR analyses provides an anchor point for how $Q_{cmb}$ has changed through time, and is thus a crucial parameter for reconstructions of the evolution of the lunar orbit and Earth rotation \cite<e.g.>{williams01,cuk16,cuk19}.  Since the lunar rotation was faster in the past, and the misalignment between the spin vectors of the mantle and core also larger, $Q_{cmb}$ was larger in the past, possibly sufficiently large to power an ancient lunar dynamo \cite{williams01,dwyer11,cebron19}.  If a part of the present-day dissipation is due instead to viscous friction at the ICB, this may impact the conclusions of these studies, and opens the possibility that the frictional heat released at the ICB may have driven convective flows with sufficient vigour to generate a dynamo \cite{stys20}.

The goal of our study is to investigate the relative contributions from friction at the CMB and ICB to the observed rotational energy dissipation of the Moon.  Whether the Moon has a solid inner core remains unknown, although its presence is expected from thermal evolution models \cite<e.g.>[]{laneuville14,zhang13,scheinberg15} and has been suggested from seismic data \cite{weber11} and  inversions of geodetic observations \cite<e.g.>[]{matsumoto15,matsuyama16}.  Viscous friction at the ICB depends on the misalignment between the spin axes of the fluid core and inner core.  The tilt angle of the spin-symmetry axis of the inner core is set by the frequency of the free inner core nutation (FICN) (DW16, SD18).  Because the FICN frequency is expected to be close to the precession frequency $\Omega_p$, the Cassini state of the Moon may feature a relatively large inner core tilt as a result of resonant amplification (DW16, SD18). The differential angular velocity at the ICB may then be larger than at the CMB.  The FICN frequency, in turn, depends on the interior density structure.  The relative contributions from friction at the ICB and CMB thus depend on the choice of lunar interior model.  Here, we sweep through a range of possible interior models parameterized in terms of inner core and outer core radii.

An additional motivation for our study is to revisit the suggestion made in \citeA{stys20} that an ancient lunar dynamo may have been powered by thermal convection from the heat released by viscous friction at the ICB.  In  \citeA{stys20}, the differential velocities at the CMB and ICB were computed from a rotational model that did not include dissipation. While the amplitude of the viscous coupling at the CMB was constrained to match its amplitude inferred by LLR, viscous coupling at the ICB was not; it was instead predicted based on a similar coupling parametrization than at the CMB but involving the differential velocity at the ICB. This simple approach, however, does not ensure that the added dissipation at the ICB remains consistent with the total rotational dissipation observed through the phase lead of $\phi_p = 0.27$ arecsec.  In contrast, here we seek to determine the viscous friction at both the CMB and ICB while enforcing that the observed phase lead is matched.  As we will show, adopting this self-consistent approach reduces the prediction of the viscous heating at the ICB by a few orders of magnitude.

\section{Method}

\subsection{Interior structure}

We adopt a simple model of the lunar interior that consists of four layers of uniform density: a solid inner core, a fluid outer core, a solid mantle, and a thin crust.  We set the lunar mass to $M=7.3463 \times 10^{22}$ kg and its mean outer radius to $R = 1737.151$ km.  We chose a crustal thickness of 38.5 km with a density of 2,550 kg m$^{-3}$ \cite{wieczorek13}.  The inner core density is fixed at 7,700 kg m$^{-3}$ \cite<e.g.>[]{matsuyama16}.  To cover a range of possible interior models, we sweep through an array of possible ICB and CMB radii.  We follow the procedure detailed in section 3 of OD20; for each combination of ICB and CMB radii, the density of the mantle is determined by matching the moment of inertia of the solid shell $I_{sm} = 0.393112  \cdot M R^2$ and the density of the fluid core is then determined by matching the lunar mass.

Each layer is triaxial in shape.   We assume that the ICB and CMB are both at hydrostatic equilibrium with the imposed gravitational potential from the triaxial shapes of the exterior surface and crust-mantle boundary.  The global triaxial shape is set so that it matches the degree 2 gravitational potential coefficients $J_2$ and $C_{22}$ and the observed polar ($\varepsilon_r$) and equatorial ($\xi_r$) flattenings of the exterior surface.  The triaxial shape at each interior boundary is found by the procedure detailed in section 3.1 of SD18.  The numerical values for $J_2$, $C_{22}$, $\varepsilon_r$ and $\xi_r$ are taken as those given in Table 1 of OD20.

\subsection{Rotational model of the Cassini state}

To capture the Cassini state of the Moon, we use the rotational model described in detail in OD20 and summarized here.  This model is a refined version of that presented in DW16.  It consists of a system of five equations and five unknowns.  The five unknowns are rotational variables.  They are: the angle of tilt of the lunar figure axis (or, equivalently, the axial symmetry axis) with respect to the ecliptic normal ($\tilde{p}$); the misalignment angle of the spin axis of the solid outer shell (comprised of the mantle and crust) with respect to the figure axis ($\tilde{m}$); the misalignment angles of the spin axes of the fluid core ($\tilde{m}_f$) and inner core ($\tilde{m}_s$) with respect to $\tilde{m}$; and the tilt angle of the inner core figure with respect to the lunar figure axis ($\tilde{n}_s$).  

Neglecting small amplitude librations, these angles are fixed when viewed in a frame attached to the Cassini plane.  Figure 2 of OD20 shows a graphical representation of these angles (with labels $\theta_p$, $\theta_m$, $\theta_f$, $\theta_s$ and $\theta_n$ corresponding respectively to $\tilde{p}$, $\tilde{m}$, $\tilde{m}_f$, $\tilde{m}_s$ and $\tilde{n}_s$).  Forced by the precession of the lunar orbit, the orientation of the Cassini plane is rotating in a retrograde direction in inertial space at frequency $\Omega_p=2\pi/18.6$ year$^{-1}$.  The equations of the rotation model of OD20 are developed in a frame attached to the mantle and crust rotating at sidereal frequency $\Omega_o=2\pi/27.322$ day$^{-1}$.  Viewed in the mantle frame, the Cassini plane is then rotating in a retrograde direction at frequency ${\omega} \Omega_o = - 2\pi/27.212$ day$^{-1}$, where ${\omega}$, expressed in cycles per lunar day, is equal to 

\begin{equation}
{\omega} = -1 - \delta {\omega} \, .
\label{eq:omega1}
\end{equation}
The factor $\delta {\omega} = \Omega_p/\Omega_o = 4.022 \times 10^{-3}$ is the Poincar\'e number, expressing the ratio of the forced precession to sidereal rotation frequencies.  The frequency ${\omega} \Omega_o$ captures the time it takes for the Moon to return to the same nodal point in its orbit.  It is the forcing frequency associated with the Cassini state when viewed in the rotating lunar frame, and the leading order tidal frequency acting on the Moon.
 
The tilde notation used for the rotational variables expresses a complex amplitude, with the real and imaginary parts capturing respectively the tilt angle components that are parallel and orthogonal to the Cassini plane. Out-of-plane (imaginary) components result from dissipation mechanisms.  As they are fixed to the Cassini plane, when viewed in the mantle frame, the rotational variables execute a retrograde precession at frequency ${\omega} \Omega_o$.  Their time-dependent part is expressed by $\exp[{i {\omega} \Omega_o t}]$, where $i=\sqrt{-1}$ is the imaginary number.  

The five equations of the model are given in Equation (54) of OD20.  The first three capture the rate of change of the angular momenta of the whole Moon, the fluid core, and the inner core, respectively.  These include the external gravitational torque from the Earth acting on the figures of the moon and its inner core, the gravitational and pressure torques between each regions, and  the torques from viscous friction at the CMB and ICB.   The last two equations of the model are kinematic relations, one to express the change in the orientation of the inner core figure resulting from its own rotation and a second describing the invariance of the ecliptic normal as seen in the mantle frame.  Each of these equations are developed under the assumption that the five unknown angles are small, and the system of equations can be expressed in a compact form as 

\begin{equation}
\boldsymbol{\mathsf{M}} \cdot {\bf x} = {\bf y} \, ,\label{eq:mat}
\end{equation}
with solution vector ${\bf x} = [ \tilde{m},  \tilde{m}_f,  \tilde{m}_s,  \tilde{n}_s, \tilde{p} ]^T$.  The elements of the matrix $\boldsymbol{\mathsf{M}}$ and right-hand side vector ${\bf y}$ are given in Appendix A of OD20.

For a given interior density structure and triaxial figure of the Moon, the rotational model captures the angular momentum response of the Moon when submitted to the external gravitational torque and tidal deformation by Earth at frequency $\omega \Omega_o$.  The solution, the mutual alignment of the five rotational variables, is the Cassini state of the Moon.  To match observations, a successful model of the Cassini state should then predict a tilt of the figure axis of $\theta_p=1.543^\circ$ (i.e. $Re[\tilde{p}]= 1.543^\circ$) and a phase lead of $\phi_p=0.27$ arcsec (i.e. $Im[\tilde{p}] = -\phi_p = -0.27$ arcsec; note the negative sign, a phase lead corresponds to a negative imaginary part.)

The objective of OD20 was to investigate the possible contribution from viscous relaxation within the inner core to the rotational dissipation.  Viscous friction at the  CMB and ICB, though included as part of the model construction, were turned off.  Here, we proceed in reverse: we assume no viscous relaxation within the inner core, and explore how viscous friction at both the CMB and ICB are connected to the observed $\phi_p$.  The model of viscous friction at the CMB and ICB is presented in section 2.4.

One important aspect of the model to note is that although the triaxial shape of the Moon is used in the prescription of the gravitational torque from Earth, the angular momentum response is based on an axisymmetric body.  To first order this is correct as the rotational response is determined by the resonant amplification of three free modes of rotation (the free precession, the free core nutation (FCN) and the FICN) which are quasi-circular motions even for a triaxial body \cite<e.g.>[]{peale05,vanhoolst02}. The convenience of adopting such a framework is that, for each region, the two equatorial angular momentum equations can be combined into a single equation.  

Also note that the flow motion in the fluid core is oversimplified in our model.  The only flow component that is explicitly tracked is its solid body rotation. The justification validating this approach are presented in \citeA{mathews91a} from which the model of OD20 is adapted.  Nevertheless, this implies that possible dynamical contributions may be missing in our model, including inertial waves, which can interact with and alter the FCN and FICN precession modes \cite<e.g.>[]{rogister09,triana19,rekier20}.  Likewise, our model does not take into account possible non-linear interactions between core waves that may have a feedback on its rotation at monthly period.

\subsection{Viscoelastic deformations from solid body tides}

Viscoelastic deformations in the rotational model of OD20 are captured by perturbations in the moments of inertia of the inner core, fluid core and the whole Moon.  These perturbations, in turn, are parameterized by a set of compliances $S_{ij}$.  The perturbations are split into an internal contribution -- from the changes in the centrifugal and gravitational potentials induced by the misaligned orientations of each layer -- and an external contribution -- from the gravitational potential of Earth.  The latter results in solid body tides of harmonic degree 2.  The part of the lunar deformation that is in-phase with the imposed external potential (the elastic part of the deformation) is captured by the tidal Love number $k_2$.  The out-of-phase component, indicative of viscous or anelastic deformation and tied to  dissipation, is captured by a quality factor $Q$, with $Q^{-1}$ representing the fraction of the total energy that is dissipated over one cycle.  A low (high) Q value indicates a high (low) dissipation.  These are connected to the compliance $S_{11}$ through 

\begin{equation}
    Re[S_{11}] = k_2 \frac{R^5\Omega_0^2}{3G\bar{A}} \, , \hspace{1cm}     Im[S_{11}] = \frac{k_2}{Q} \frac{R^5\Omega_0^2}{3G\bar{A}} \, ,\label{eq:k2Q}
\end{equation}
where $G$ is the gravitational constant, and $\bar{A}$ is the mean equatorial moment of inertia of the whole Moon.  

Note that $k_2$ and $Q$ (and thus $S_{11}$) capture the bulk deformation of the Moon, without giving direct information on where in its interior deformations may be maximized.  At the monthly tidal period of 27.212 days, recent observations suggest $k_2 = 0.02422\pm 0.00022$ \cite{williams14} and $k_2/Q= (6.4 \pm 1.5) \times 10^{-4}$ \cite{williams15} corresponding to a monthly Q-value of 37.8.   Interior models that include a low viscosity zone in the lowermost mantle  \cite<e.g. OD20,>[]{harada14,harada16}, possibly featuring partial melt \cite<e.g.>[]{khan14}, are consistent with these $k_2$ and $Q$ values.  If this is correct, tidal dissipation is concentrated in the lowermost region of the mantle.

To a very good approximation, the tidal contribution to the phase lead $\phi_p$ in the rotational model of OD20 is determined by the imaginary part of $S_{11}$.  Hence, we set all compliances to zero, except $S_{11}$.  We do not compute $S_{11}$ from a model of seismic parameters and viscosity within each layer, as was done in OD20, but instead we constrain it to match the central values of $k_2=0.02422$ and $k_2/Q= 6.4 \times 10^{-4}$ quoted above through Equation (\ref{eq:k2Q}).  With these choices, tides contribute approximately 0.15 arcsec of the observed $\phi_p$ (see section 4.2 of OD20 and our results below), with a weak dependence on the choice of inner core and outer core radii.  The remaining $\sim 0.12$ arcsec required to match the observed $\phi_p$ must then be accommodated by viscous friction at the CMB and ICB.

\subsection{Viscous torque at the CMB and ICB}

The torques from viscous friction at the CMB ($\tilde{\Gamma}_{cmb}$) and ICB ($\tilde{\Gamma}_{icb}$) are parameterized as products between dimensionless complex coupling constants ($K_{cmb}$ and $K_{icb}$) and the differential angular velocities at each boundary.   In the complex notation used in OD20, they are given by their Equation (52), 

\begin{subequations}
\begin{equation}
    \tilde{\Gamma}_{cmb} = i \Omega_o^2 \bar{A}_f K_{cmb} \, \tilde{m}_f \, ,\label{eq:tqcmb}
\end{equation}
\begin{equation}
    \tilde{\Gamma}_{icb} = i \Omega_o^2 \bar{A}_s K_{icb} (\tilde{m}_f - \tilde{m}_s) \, ,\label{eq:tqicb}
\end{equation}
\label{eq:tqvisc}
\end{subequations}
where $\bar{A}_f$ and $\bar{A}_s$ are the mean equatorial moments of inertia of the fluid core and inner core, respectively.  Expressions for $K_{cmb}$ and $K_{icb}$ depend on whether the flow in the viscous boundary layer (Ekman layer) remains stable (i.e. laminar) or not (turbulent). 

At the CMB of the Moon (radius $r_f$), the differential velocity between the mantle and fluid core, ${\cal U} = r_f \Omega_o | \sin(\tilde{m}_f)|$, is sufficiently large to induce instabilities \cite{toomre66,yoder81,williams01,cebron19}. The stability of the Ekman layer is determined by the local Reynolds number $R_e = {\cal U} \delta/\nu$, where  $\nu$ is the kinematic viscosity and $\delta = \sqrt{\nu/\Omega_o}$ is the Ekman layer thickness.  For an oscillating differential velocity, the boundary layer is expected to be in a turbulent regime when $R_e > 500$ \cite<e.g.>{buffett21}. With $\Omega_o=2.6617\times10^{-6}$ s$^{-1}$, and taking $r_f=380$ km \cite<e.g.>{viswanathan19} and $\tilde{m}_f=-1.6^\circ$ (e.g. SD18) gives ${\cal U} = 2.82$ cm s$^{-1}$.  The kinematic viscosity of liquid iron in planetary cores is expected to be of the order of $10^{-6}$ m$^2$ s$^{-1}$ \cite<e.g.>{alfe00,rutter02b,rutter02a}, which gives $\delta = 61.3$ cm and $R_e = 1.73\times 10^4$.  This is far above the threshold $R_e > 500$ and the flow in the Ekman layer at the CMB is expected to be in a turbulent regime.  

Likewise, the flow in the Ekman layer at the ICB is also likely in a turbulent regime.  The misalignment between the rotation vectors of the fluid and solid cores is highly sensitive to the choice of interior model (DW16, SD18), but taking $|\tilde{m}_f - \tilde{m}_s| = 4^\circ$ as a representative measure (see our results below), the differential velocity at the ICB (radius $r_s$), ${\cal U} = r_s \Omega_o | \sin(\tilde{m}_f - \tilde{m}_s)|$, is sufficiently large that $R_e$ is above the threshold of $500$ as long as the for an inner core radius is larger than 4.4 km.

For a turbulent flow, the viscous shear stress on the solid boundary is written as $\bm{\tau} = \kappa \rho_f |{\bf u}| {\bf u}$, where $\rho_f$ is the fluid core density, ${\bf u}$ is the flow velocity outside the boundary layer and $\kappa$ is a drag coefficient that depends on viscosity, rotation, $|{\bf u}|$ and surface roughness \cite<e.g.>[]{sous13}.  Integrating ${\bf r} \times \bm{\tau}$ over the spherical surfaces of the CMB and ICB, where  ${\bf r}$ is the radial vector, and assuming $\kappa$ is uniform, we can write the viscous torques in the form of Equations (\ref{eq:tqvisc}) with the coupling constants $K_{cmb}$ and $K_{icb}$ given by 

\begin{subequations}
\begin{equation}
    K_{cmb} = -i \frac{3 \pi^2}{4} \kappa_{cmb}\left|\tilde{m}_f\right|\, , \label{eq:turbcmb}
\end{equation}
\begin{equation}
    K_{icb} = -i \frac{3 \pi^2}{4} \frac{\rho_f}{\rho_s}  \kappa_{icb}\left|\tilde{m}_s-\tilde{m}_f\right| \, ,\label{eq:turbicb}
\end{equation}
\label{eq:Kcmbicb}
\end{subequations}
where $\kappa_{cmb}$ and $\kappa_{icb}$ denote the drag coefficients at the CMB and ICB, respectively.  The factor $\rho_f/\rho_s$ in $K_{icb}$ accounts for the fact that it is the density of the fluid core which is involved in the viscous stress at the ICB.  Our expression for $K_{cmb}$ matches that used by \citeA{williams01} and \citeA{cebron19}.  

The numerical values of the drag coefficients $\kappa_{cmb}$ and $\kappa_{icb}$ are a priori unknown.  We can either prescribe specific values and monitor the consequence of these choices on the solution of our rotational model.  We proceed in this manner for the results that are presented in Figure \ref{fig:sol} and described in Section 3.  The alternative is to search for a set of $\kappa_{cmb}$ and $\kappa_{icb}$ that allows us to match $\phi_p=0.27$ arcsec and therefore be consistent with the observed dissipation.  This is the approach that we take for the results presented in Figures \ref{fig:rsrf1} and \ref{fig:rsrf2}.  For simplicity, we assume in all our calculations that $\kappa_{cmb}=\kappa_{icb}$.  
   
Before closing this section, let us add a quick note on the computation of our solutions.  $K_{cmb}$ and $K_{icb}$ enter some elements of matrix $\boldsymbol{\mathsf{M}}$ in Equation (\ref{eq:mat}) (see Appendix A of OD20).  With the turbulent model parametrization of Equation (\ref{eq:Kcmbicb}), $K_{cmb}$ and $K_{icb}$ depend on $\tilde{m}_f$ and $\tilde{m}_s$, and so the rotational model is no longer linear in the rotational variables (i.e. the matrix $\boldsymbol{\mathsf{M}}$ is itself dependent on $\tilde{m}_f$ and $\tilde{m}_s$).   For a given choice of $\kappa_{cmb}$ ($=\kappa_{icb}$), solutions are found by a fixed-point iteration method, though when convergence has not been reached after a few iterations we switch to a multi-directional Newton method.   For the cases that involve searching for the numerical value of the $\kappa_{cmb}$ ($=\kappa_{icb}$) that matches the observed $\phi_p=0.27$ arcsec, we use a one-dimensional Newton method.  When the FICN frequency is very close to $\Omega_p$, this strategy fails in some cases; when this occurs, we find $\kappa_{cmb}$ either by a bisection method or by interpolation.

\subsection{Periodic gravity signal induced by a tilted inner core}

In the Cassini state equilibrium, the spin-symmetry axis of the inner core is misaligned from that of the mantle (DW16, SD18). Viewed in the reference frame of the rotating mantle, a tilted inner core undergoes a retrograde precession with a period of 27.212 day.  This induces a periodic variation in the degree 2, order 1 coefficients of gravity \cite{williams07}.  However, such a gravity signal has not been detected to date \cite<e.g.>{williams15b}.  We use this lack of detection as an additional constraint on our rotational and interior structure models.

The periodic change in the degree 2, order 1 component of the gravity field is captured by perturbations in Stokes coefficients $C_{21}$ and $S_{21}$, and we denote their amplitudes by $|\Delta C_{21}|$ and $|\Delta S_{21}|$.   For an axially symmetric inner core, 

\begin{equation}
|\Delta  C_{21} | =|\Delta S_{21}|=\frac{ \bar{A}_s \alpha_3 e_s }{M R^2} \cos (|\tilde{n}_s|)  \sin (| \tilde{n}_s|) \, , \label{eq:deltaC21}
\end{equation}
where $\alpha_3 = 1 - \rho_f/\rho_s$, $e_s$ is the dynamical ellipticity of the inner core (Equation 6 of OD20) and $|\tilde{n}_s|$ is the magnitude of the tilt of the inner core with respect to the mantle symmetry axis. The amplitude of the periodic degree 2, order 1 gravity coefficient possibly attributable to an inner core based on GRAIL data is the range of $4-8 \times 10^{-11}$, but deviations of the order of the uncertainties ($\sim 5-7 \times 10^{-11}$) of their central values are necessary in order to match predictions \cite{williams15b}.  The absence of a clear periodic signal emerging above the noise level indicates that $|\Delta  C_{21} |$ must be below a detection baseline, which we take to be $5 \times 10^{-11}$.

Viscoelastic deformations in response to a tilted inner core and from the precessing spin vector of the fluid core, alter the prediction of Equation (\ref{eq:deltaC21}).  Furthermore, a triaxial inner core introduces a difference between $|\Delta C_{21}|$ and $|\Delta S_{21}|$ \cite<see>{williams07}.  Equation (\ref{eq:deltaC21}) is the mean of these two amplitudes and gives a simple first order guideline for the amplitude of inner core gravity signal.  We use $|\Delta  C_{21} | \leq 5 \times 10^{-11}$ to delimit the range of acceptable lunar models.  

\section{Results}

\begin{figure}
\begin{center}
    \includegraphics[width=15cm]{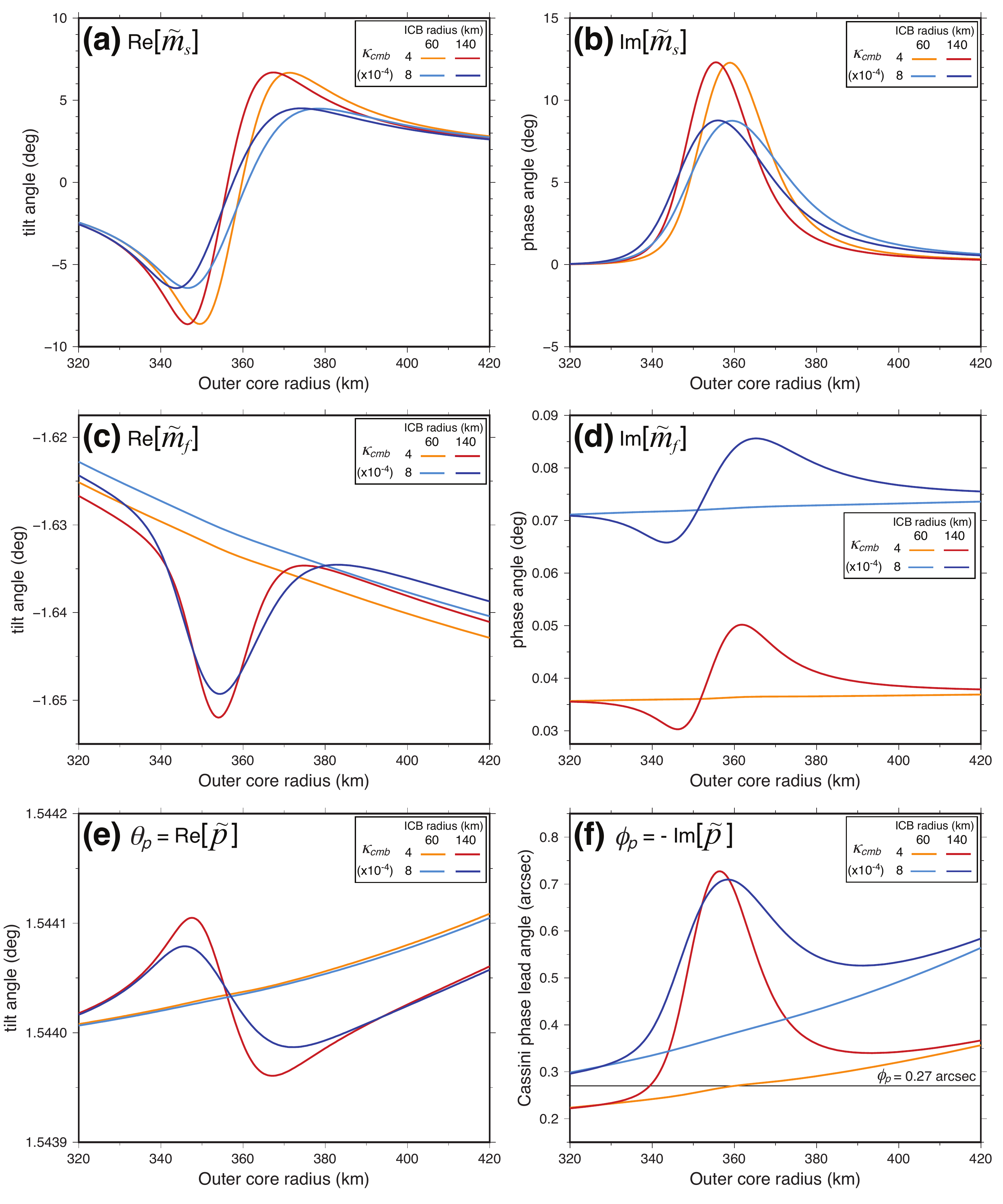}  
    \caption{\label{fig:sol} (a) $Re[\tilde{m}_s]$, (b) $Im[\tilde{m}_s]$, (c) $Re[\tilde{m}_f]$, (d) $Im[\tilde{m}_f]$, (e) $Re[\tilde{p}]$ and (f) $\phi_p=-Im[\tilde{p}]$ as a function of outer core radii for two different choices of inner core radius (60 km and 140 km) and two different choices of drag coefficient $\kappa_{cmb} = \kappa_{icb}$ ($4 \times 10^{-4}$ and $8 \times 10^{-4}$). The horizontal black line in (f) shows the observed phase lead of $\phi_p=0.27$ arcsec.}
\end{center}
\end{figure}

Let us first show examples of solutions from our rotational model for fixed choices of the drag coefficient $\kappa_{cmb}$ (=$\kappa_{icb}$).  Figure \ref{fig:sol} shows the real and imaginary parts of $\tilde{m}_s$, $\tilde{m}_f$ and $\tilde{p}$ for a range of outer core radii ($r_f$ between 320 and 420 km), two different choices of inner core radius ($r_s=60$ km and 140 km) and two different choices of $\kappa_{cmb}$ ($4 \times 10^{-4}$ and $8 \times 10^{-4}$).  The form of the solutions for $\tilde{m}$ (not shown) is identical to that of $\tilde{p}$ since they are connected by $\tilde{m}= - (1+\omega) \tilde{p} = \delta \omega \, \tilde{p}$ (see Equation 54e of OD21), though $\tilde{m}$ has a much smaller magnitude.   The solutions for $\tilde{n}_s$ are virtually identical to those of $\tilde{m}_s$ since $\tilde{n}_s=-\tilde{m}_s/\omega=\tilde{m}_s/(1+\delta \omega) \approx \tilde{m}_s$ (see Equation 54d of OD21); the figure and spin axes of the inner core are aligned in the Cassini state.

The transition in $Re[\tilde{m}_s]$ from negative to positive values at $r_f\sim 360$ km (Figure \ref{fig:sol}a) accompanied by a peak in $Im[\tilde{m}_s]$ (Figure \ref{fig:sol}b) marks the location in parameter space where the frequency of the FICN ($\Omega_{ficn}$) is close to the orbital precession frequency $\Omega_p= 2\pi/18.6$ yr$^{-1}$.  The FICN is a free precession of the inner core with respect to other regions of the Moon, so when $\Omega_{ficn}$ approaches $\Omega_p$, a resonant amplification of the inner core tilt  occurs (DW16, SD18).  The frequency of the FICN depends on the interior density structure, notably on the density contrast at the ICB.  In the way that our interior models are constructed, a change in CMB radius alters the density of the fluid core  (in order to conserve mass), so $\Omega_{ficn}$ changes with $r_f$ in the plots of Figure \ref{fig:sol}.  For small $r_f$, $\Omega_{ficn}<\Omega_p$, while for large $r_f$, $\Omega_{ficn}>\Omega_p$.  The specific value of $r_f$ at which $\Omega_{ficn}=\Omega_p$ depends on the inner core size, as the choice of the latter affects the fluid core density in our interior models. 

Without viscous friction at the ICB, $Re[\tilde{m}_s]$ would diverge to $\pm \infty$ on either side of the resonance crossing (see for example Fig 3 of SD18) and $Im[\tilde{m}_s]$ would be identically zero.  Adding viscous drag at the ICB keeps $Re[\tilde{m}_s]$ finite and introduces a non-zero $Im[\tilde{m}_s]$.  The latter has a positive sign, so the spin axis of the inner core  lags behind the Cassini plane.  The larger the drag coefficient $\kappa_{icb}$, the higher the viscous friction at the ICB and the more attenuated is the response of the inner core. This is observed on Figure \ref{fig:sol}ab: the amplitude of $\tilde{m}_s$ is smaller for the largest choice of $\kappa_{icb}$.   

The solutions of $\tilde{m}_f$ (Figure \ref{fig:sol}cd) show how the orientation of the spin axis of the fluid core is affected by viscous coupling at both the CMB and ICB.  $Re[\tilde{m}_f]$ is contained between -1.65$^\circ$ and -1.62$^\circ$, and is dominantly controlled by the pressure torque at the CMB caused by the misalignment between the fluid core spin axis and the elliptical shape of the CMB.  The torque from viscous friction is much smaller in magnitude. It leads to a small modification of $Re[\tilde{m}_f]$ and to a positive $Im[\tilde{m}_f$] (the spin axis of the fluid core lags behind the Cassini plane), the latter increasing in magnitude with a larger $\kappa_{cmb}$.  For a small inner core, the FICN resonance does not have a visible effect on $\tilde{m}_f$.  This is because the viscous torque acting on the fluid core at the ICB is proportional to the moment of inertia of the inner core (see Equation \ref{eq:tqvisc}) and thus to $r_s^5$; for a small $r_s$, this torque  is then much weaker than the viscous torque at the CMB.  For a large inner core, the viscous torque at the ICB is no longer negligible, and $\tilde{m}_f$ is altered by the FICN resonance.  Furthermore, the inner core and mantle are coupled by a gravitational torque, with an amplitude also proportional to the moment of inertia of the inner core.  For a large inner core, the orientation of the mantle figure is then affected by the FICN resonance through this gravitational torque, and in turn, this affects the spin axis of the fluid core through viscous coupling at the CMB.

The change in mantle orientation caused by the FICN resonance is visible on the plot of $Re[\tilde{p}]$ for $r_s=140$ km  (Figure \ref{fig:sol}e), though it represents a very small perturbation of the order of $0.0001^\circ$.  $Re[\tilde{p}]$ -- the tilt of the mantle figure with respect to the ecliptic normal -- is dominantly controlled by the external gravitational torque from Earth.  Note that for all cases shown, our solution for $Re[\tilde{p}]$ is approximately equal to $1.544^\circ$, so our model recovers to within $0.001^\circ$ the observed tilt of $\theta_p=1.543^\circ$.  This is an adequate fit given the simplifications that enter our rotational model (axial symmetry, small angles, linearization, etc.).

Finally, Figure \ref{fig:sol}f shows the prediction of the phase lead angle $\phi_p$ of the mantle ahead of the Cassini plane.  A good approximation of $\phi_p$ predicted by our rotational model is (see Appendix A)

\begin{equation}
\phi_p =  \left( \frac{1}{\delta \omega (1+e+ \delta \omega) - \beta \Phi_2} \right) \Bigg[ \left( \frac{k_2}{Q} \right) \frac{R^5\,  \Omega_o^2 \, \Phi^t }{3 G \bar{A}}  + \delta \omega\left(  \frac{\bar{A}_f}{\bar{A}} Im[\tilde{m}_f] + \frac{\bar{A}_s}{\bar{A}} Im[\tilde{m}_s] \right) - \frac{\bar{A}_s}{\bar{A}} \alpha_3 \beta_s \Phi_2 Im[\tilde{n}_s] \Bigg] \, , \label{eq:varphip_pred1}
\end{equation}  
where $\Phi_2=1.4646$ and $\Phi^t=0.5453$ are parameters involved in the gravitational torque from Earth and where $e$, $\beta$ and $\beta_s$ are dynamical ellipticities defined by Equations (6) and (31) of OD20.  The first term in the square bracket captures the contribution from tidal dissipation.  With the choice of $k_2/Q=6.4 \times10^{-4}$, tidal dissipation accounts for $0.1483$ arcsec of the total $\phi_p$ (with a very weak dependence on $r_f$ and $r_s$, variations are of the order of $10^{-4}$).  The second term (proportional to $\delta \omega$) involves the angular momentum components of the fluid core and inner core perpendicular to the Cassini plane.  These result from viscous coupling at the ICB and CMB, so this second term captures the contribution to $\phi_p$ from viscous dissipation within the core.   The third and last term is a small additional correction to $\phi_p$. It captures the analog of a tidal dissipation associated with the tilted figure of the inner core; although the inner core in our models does not deform, the non-zero $Im[\tilde{n}_s]$ induced by the viscous torque at the ICB mimics a delayed tidal deformation on which the gravitational torque from Earth acts.  Equation (\ref{eq:varphip_pred1}) is based on an angular momentum balance, so it does not contain any direct information on the nature of the torques acting on the mantle and causing its phase lead.  The torque by the fluid core is from viscous friction at the CMB; the torque by the inner core is from gravitational coupling. 

For the small inner core cases shown in Figure \ref{fig:sol}, $\bar{A}_s \ll \bar{A}_f$ and the contribution to $\phi_p$ from the terms that involve $\tilde{m}_s$  and $\tilde{n}_s$ in Equation (\ref{eq:varphip_pred1}) is negligible.  In other words, with a small inner core, dissipation in the core is dominated by viscous friction at the CMB. The amplitude of the CMB viscous torque, and thus $\phi_p$, increases with $\kappa_{cmb}$ and with CMB radius.   With the combination of $\kappa_{cmb}=0.0004$, $r_s = 60$ km and $r_f = 360$ km, the sum of tidal and viscous dissipation at the CMB reproduces the observed phase lead of $\phi_p=0.27$ arcsec.  For a large inner core, the terms involving $\tilde{m}_s$ and $\tilde{n}_s$ are no longer negligible in Equation (\ref{eq:varphip_pred1}).  The amplitude of $\phi_p$ increases in the vicinity of the FICN resonance, mirroring the bump observed in $Im[\tilde{m}_s]$.

Figure \ref{fig:sol}f illustrates how, for a large inner core, the added dissipation from viscous friction at the ICB can increase the phase lead angle $\phi_p$.  The increase in $\phi_p$ is not a simple function of inner core size; rather, it is maximized when $\Omega_{ficn}$ approaches $\Omega_p$.  With the added dissipation at the ICB, the value of the drag coefficient $\kappa_{cmb}$ (=$\kappa_{icb}$) that allows to match the observed phase lead of $\phi_p=0.27$ arcsec must be reduced, the more so the closer $\Omega_{ficn}$ is to $\Omega_p$.  

Figure \ref{fig:sol}f further illustrates how, in principle, constraints on the CMB and ICB radii can be extracted from the observed $\phi_p=0.27$ arcsec.   This requires an accurate theoretical model of the turbulent viscous torque, including how the drag coefficient depends on the differential precession velocity at each boundaries.  Provided such a model is available, we can illustrate how this may work.  Let us suppose that the form of the turbulent viscous torque in Equations (\ref{eq:tqvisc}-\ref{eq:Kcmbicb}) is correct and that a theoretical model would predict a drag coefficient of $\kappa_{cmb}=0.0004$.  If the inner core is small ($r_s < 60$ km), then based on Figure \ref{fig:sol}f, the radius of the CMB would then be approximately 360 km.  For $r_s=140$ km (and assuming $\kappa_{icb}=\kappa_{cmb}$), the radius of the CMB would be slightly smaller, approximately 340 km.  The error on these estimates depends on the combination of the uncertainties on $\kappa_{cmb}$, $\kappa_{icb}$ and $k_2/Q$ (which sets the amount of tidal dissipation).  At present, these errors remain too large to extract meaningful constraints on $r_f$ and $r_s$.  To illustrate this, the $k_2/Q$ parameter inferred from LLR is $(6.4 \pm 1.5)\times 10^{-4}$, so taking its error into account, it maps to a tidal contribution to $\phi_p$ of $0.1483 \pm 0.0348$ arcsec.  For $r_s=60$ km and $\kappa_{cmb}=0.0004$, matching $\phi_p=0.27$ arcsec with the addition of viscous coupling maps to a range of possible CMB radii between 334 and 390 km.  Hence, even if $\kappa_{cmb}$ could be accurately predicted (which is not the case, a point we return to in the discussion), the current error on $k_2/Q$ is too large to significantly narrow down the range of possible interior lunar models.  

\begin{figure}
\begin{center}
     \includegraphics[width=15cm]{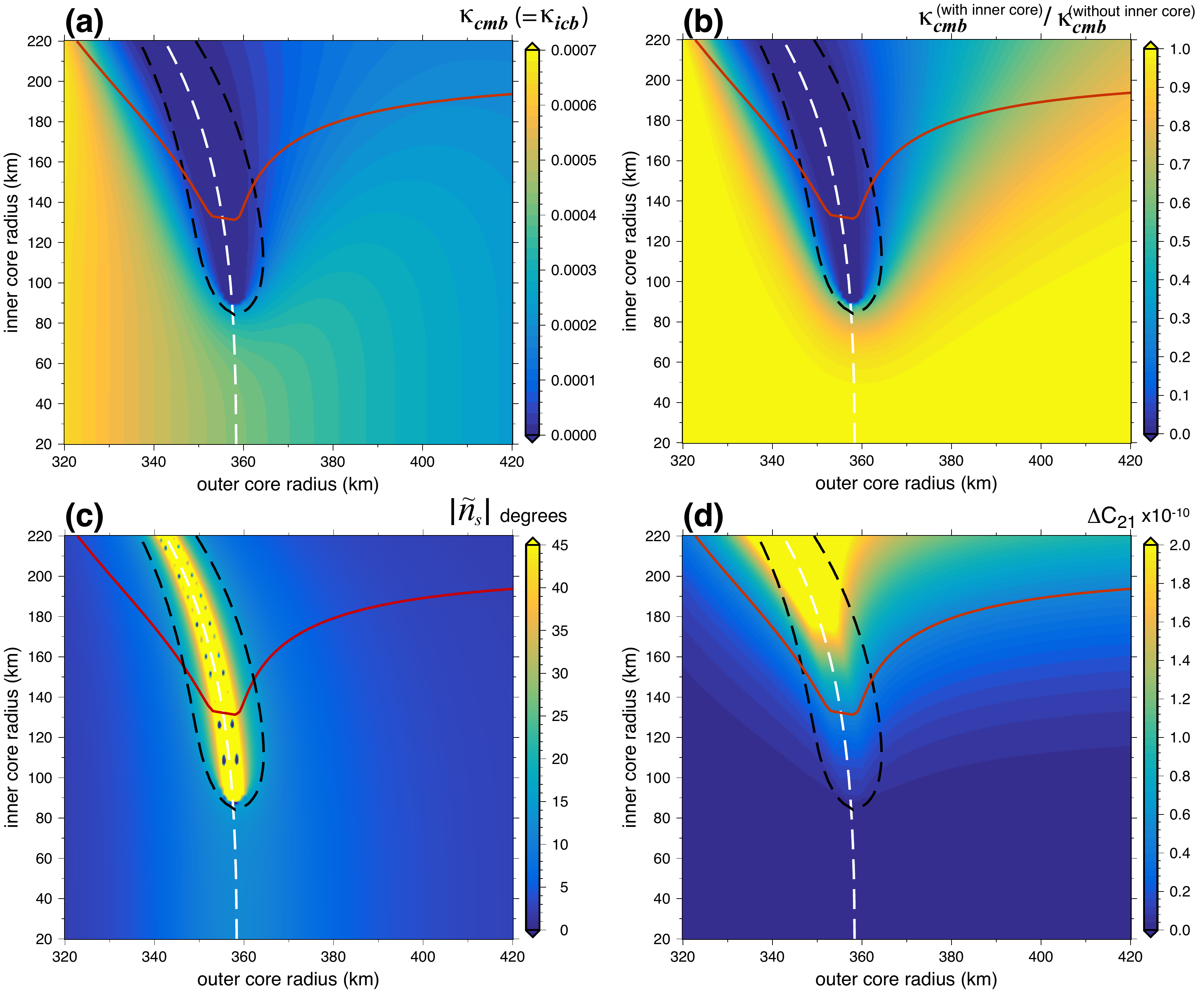} 
\caption{\label{fig:rsrf1} (a) The numerical value of the drag coefficient $\kappa_{cmb}$, (b) the ratio of $\kappa_{cmb}$ with vs without an inner core, (c) the magnitude of the inner core tilt $ | \tilde{n}_s |$ with respect to the mantle, and (d) the amplitude of the periodic degree 2 order 1 gravity signal $ | \Delta C_{21} |$ associated with a precessing inner core, as a function of outer core and inner core radii such that $\phi_p=0.27$ arcsec.  The white dashed line marks where the FICN frequency is equal to the orbital precession frequency.  The solid red contour line corresponds to $ | \Delta C_{21} |=5\times10^{-11} $.  The black dashed contour line  indicates  where $ | \tilde{n}_s |=15^\circ$.  The colour contours in (c) and (d) are saturated at $45^\circ$ and $2\times10^{-11}$, respectively.}
    \end{center}
\end{figure}

With a given choice of $\kappa_{cmb}$, only specific combinations of ICB and CMB radii can match the observed phase lead of $\phi_p=0.27$ arcsec.  Conversely then, for a given combination of ICB and CMB radii, only a specific value of $\kappa_{cmb}$ permits to match $\phi_p=0.27$ arcsec.  Figure \ref{fig:rsrf1}a shows how $\kappa_{cmb}$ must be adjusted as a function of $r_s$ and $r_f$ such that the combination of tidal dissipation and viscous friction at both the ICB and CMB results in a phase lead of $\phi_p=0.27$ arcsec. Figure \ref{fig:rsrf1}b  shows the fractional change of $\kappa_{cmb}$ compared to that computed in the absence of an inner core.  $\kappa_{cmb}$ must be smaller than $0.0001$ if the inner core radius is larger than approximately 85 km and when the $\Omega_{ficn}$ approaches $\Omega_p$ (the combination of $r_f$-$r_s$ for which $\Omega_{ficn}=\Omega_p$ is indicated by the white dashed line). 

If a theoretical model of $\kappa_{cmb}$ predicts a specific value, (say $0.0004$, to continue the example above), then this contour line on Figure \ref{fig:rsrf1}a delineates the combinations of ICB and CMB radii that are compatible with the observed dissipation (and under the assumption that $k_2/Q = 6.4 \times 10^{-4}$).  Multiple combinations of ICB and CMB radii remain possible, but Figure \ref{fig:rsrf1}a illustrates nevertheless how in principle the core geometry could be further constrained.  For example,  for a specific CMB radius, the range of possible ICB values on Figure \ref{fig:rsrf1}a should be restricted to those that fall within the range of $\kappa_{cmb}$ values that are consistent with theoretical predictions.  Redoing this exercise for the upper and lower bounds of $k_2/Q$ allowed by its error bar would provide the complete range of possible ICB radii.

Figure \ref{fig:rsrf1}c shows the magnitude of the inner core tilt angle ($|\tilde{n}_s|$) with respect to the mantle symmetry axis for the same set of solutions.  When $|\tilde{n}_s|$ is larger than approximately $15^\circ$ (marked by the black dashed contour line on all panels of Figure \ref{fig:rsrf1}), the small angle assumption of our rotational model is no longer valid and the results are no longer accurate.   As shown in SD18 with a rotational model that is not limited to small angles (though without dissipation), the component of the inner core tilt in the Cassini plane is restricted to the range $[-33^\circ, +17^\circ]$.  Predictions from our linear model that significantly exceed this are then unrealistically large.  By inspection of Figure \ref{fig:rsrf1}a, the lunar models for which $\kappa_{cmb}$ approaches zero correspond to cases with unrealistically large inner core tilts in excess of 20$^\circ$; vanishingly small values of $\kappa_{cmb}$ are then an artefact due to the limitation of our rotational model.  The solutions associated with the $|\tilde{n}_s|=15^\circ$ contour line, though not accurate, give a reasonable indication of the solutions that can be expected in a more accurate model.

The prediction of the degree 2, order 1 gravity signal $| \Delta C_{21}|$ is shown in Figure \ref{fig:rsrf1}d.  The red contour line marks the detection baseline $| \Delta C_{21}| = 5 \times 10^{-11}$ above which the periodic gravity signal associated with the inner core tilt should have been detected.  This contour line is also drawn on the other panels of Figure \ref{fig:rsrf1}.  Lunar models that have $| \Delta C_{21}| > 5 \times 10^{-11}$ are inconsistent with the non-detection of this gravity signal.  Acceptable lunar models are restricted to those with an inner core size which falls below this demarcation line.

The overall picture that emerges from Figure \ref{fig:rsrf1} is that, for models compatible with $| \Delta C_{21}| < 5 \times 10^{-11}$, in a section of the $r_f-r_s$ parameter space where the FICN frequency $\Omega_{ficn}$ is close to $\Omega_p$ (to within 30\%) and the inner core radius is larger than 80 km, the drag coefficient $\kappa_{cmb}$ must be substantially reduced (by as much as a factor 10) in order to match $\phi_p=0.27$ arcsec.  The reason is because of the added contribution to the dissipation by viscous friction at the ICB; without a reduction $\kappa_{cmb}$, $\phi_p$ exceeds 0.27 arcsec. If the inner core is smaller than 80 km, of if $\Omega_{ficn}$ departs sufficiently from $\Omega_p$, $\kappa_{cmb}$ is not substantially smaller than its value for a small or no inner core.  

\begin{figure}
\begin{center}
         \includegraphics[width=15cm]{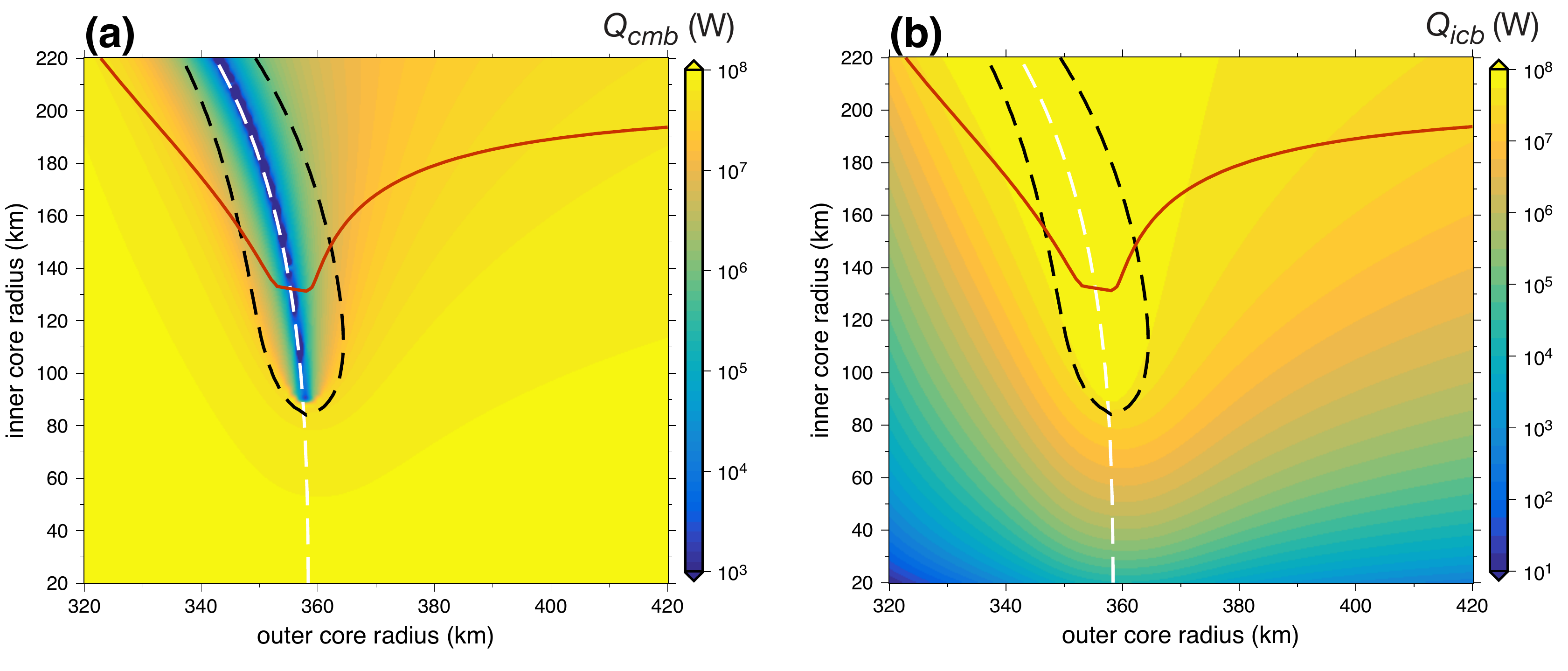} 
\caption{\label{fig:rsrf2} The viscous dissipation (in Watts) at the (a) CMB and (b) ICB, as a function of outer core and inner core radii such that $\phi_p=0.27$ arcsec.  These corresponds to the same model solutions shown in Figure \ref{fig:rsrf1}.  The white dashed line marks where the FICN frequency is equal to the orbital precession frequency.  The solid red contour line corresponds to $ | \Delta C_{21} |=5\times10^{-11} $.  The black dashed contour line  indicates  where $ | \tilde{n}_s |=15^\circ$.}
    \end{center}
\end{figure}

Figure \ref{fig:rsrf2} shows the amplitude of the viscous dissipation at the CMB ($Q_{cmb}$) and ICB ($Q_{icb}$) for the solutions shown in Figure \ref{fig:rsrf1}. These are computed from

\begin{subequations}
\begin{equation}
Q_{cmb} =\kappa_{cmb}  \frac{3\pi^2}{4}  \, {\cal I}_f \Omega_o^3 \, \big|  \tilde{m}_f \big|^3 \, , 
\end{equation}
\begin{equation}
Q_{icb} =  \kappa_{cmb} \frac{3\pi^2}{4} \frac{\rho_f}{\rho_s} {\cal I}_s \Omega_o^3 \,  \big| \tilde{m_s}-\tilde{m}_f \big|^3\, , 
\end{equation}
\end{subequations}
where ${\cal I}_s=(8\pi/15)\rho_s r_s^5$ and ${\cal I}_f=(8\pi/15)\rho_f r_f^5$ are the mean moments of inertia of the solid inner core and an entirely fluid core, respectively.  For a small inner core, $Q_{cmb}$ is approximately equal to $8.18 \times 10^7$ W.  For models that obey $ | \Delta C_{21} |\leq 5\times10^{-11} $, $Q_{cmb}$ is not substantially reduced and $Q_{icb}$ is typically smaller than $10^7$ W in the section of the parameter space where friction at the CMB still dominates.  However,  when $r_s>80$ km and $\Omega_{ficn}\approx\Omega_p$, $Q_{cmb}$ can be reduced to below $10^7$ W and $Q_{icb}$ can reach values as high as $7.70 \times 10^7$ W.  Note that in the section of parameter space where $Q_{icb}$ is high, the sum of $Q_{cmb}$ and $Q_{icb}$ is lower than $8.18 \times 10^7$ W, the total dissipation within the core in the absence of an inner core.  The latter is conserved for all models and is equal to $8.18 \times 10^7$ W; the missing part is due to the tidal-like dissipation associated with the tilted inner core. This latter part never accounts for more than 10\% of the total dissipation in the core, so in regions of the parameter space where $Q_{cmb}$ is reduced, it is dominantly because $Q_{icb}$ is increased. 

Lastly, we reiterate a point made in OD20, that all results presented in Figures \ref{fig:sol}-\ref{fig:rsrf2} are tied to the choices we have made for the density and thickness of the crust. These influence the densities of the mantle and fluid core in the way that we constrain our interior lunar models.  In turn, this affects the frequency of the FICN for a given combination of $r_s$ and $r_f$.  With different assumptions about the crust, the location of the FICN resonance would be shifted and so would the lines and contours of each quantities plotted on Figures \ref{fig:sol}-\ref{fig:rsrf2}.  The general trends as a function of $r_s$ and $r_f$ would remain unaltered, but one should be careful in extracting a specific numerical values for a choice of $r_s$ and $r_f$ from each of these Figures.

\section{Discussion and Conclusions}

Predictions of the amplitude of the periodic degree 2, order 1 gravity signal induced by a precessing inner core were presented in \citeA{williams07} and \citeA{williams15b}, although only for a few specific cases.  Figure \ref{fig:rsrf1}d complement these predictions for a range of ICB and CMB radii.  We recall that our predictions correspond to the mean of the amplitudes of the periodic gravity coefficients $C_{21}$ and $S_{21}$, or equivalently the predictions based on an axially symmetric inner core.  The non-detection of this gravity signal in the data collected by the GRAIL satellite mission \cite<e.g.>{williams15b} offers then a constraint on plausible lunar interior models, in particular the size of the inner core.  Based on a detection baseline set at $5\times 10^{-11}$, the inner core must be smaller than $\sim200$ km if $\Omega_{ficn} > \Omega_p$ (i.e. to the right of the white dashed lines on Figures \ref{fig:rsrf1}-\ref{fig:rsrf2}).  A larger maximum inner core radius is possible if the CMB radius is at the small end of our range (and thus if $\Omega_{ficn} < \Omega_p$).  The closer $\Omega_{ficn}$ is to being in resonance with $\Omega_p$, the smaller is the maximum inner core radius; for $\Omega_{ficn}=\Omega_p$, it is approximately $\sim130$ km.   The maximum inner core size allowed in Figure \ref{fig:rsrf1}d  is slightly more restrictive than that inferred from matching the lunar mass and the moment of inertia of the solid part of the Moon, which gives an upper bound of approximately 280 km \cite{williams14}.   

As our results further illustrate, additional constraints on the CMB and ICB radii can possibly be extracted from the requirement that the combination of tidal and viscous dissipation must reproduce the observed phase lead of $\phi_p = 0.27$ arcsec.   This requires an accurate determination of the tidal dissipation (through the parameter $k_2/Q$) and an accurate theoretical model of the turbulent viscous torque at the fluid core boundaries.  At present, the error on $k_2/Q$ remains too large to extract useful constraints on the core geometry.  Likewise, predictions of the turbulent viscous torque remain imprecise.  To illustrate this last point, the most recent theoretical model of the drag coefficient $\kappa_{cmb}$ appropriate for the lunar precession is that given in the study of \citeA{cebron19}.  The present-day CMB viscous dissipation based on their model is approximately a factor 2 larger than that deduced from LLR (see their Figure 24).  Although their model of $\kappa_{cmb}$ succeeds in recovering approximately the correct viscous dissipation in a turbulent regime, it remains not sufficiently accurate for the purpose of narrowing down the combinations of ICB and CMB radii that are compatible with the observed dissipation.  If models of the turbulent viscous torque improve, and provided the error bar on $k_2/Q$ can be reduced, matching the observed $\phi_p$ in a rotational model like the one we have developed can yield constraints on the lunar core.

If we restrict possible lunar models to those for which $|\Delta C_{21}| \leq 5 \times 10^{-11}$, our results show that it is only when the ICB radius $r_s$ is larger than 80 km and when $\Omega_{ficn}$ is close to $\Omega_p$ (approximately within 30\%), that viscous dissipation at the ICB  ($Q_{icb}$) becomes comparable (and may even surpass) viscous dissipation at the CMB ($Q_{cmb}$).  In such cases, $Q_{cmb}$ can be reduced by as much as a factor 10 compared to a lunar model without an inner core, and $Q_{icb}$ can reach values as high as $7.7 \times10^7$ W.  For $r_s<80$ km or if $\Omega_{ficn}$ departs from $\Omega_p$ by more than 30\%, $Q_{cmb}$ is reduced by no more than approximately 10\% compared to a lunar model without an inner core and $Q_{icb}$ is weaker than $10^7$ W.

A caveat on all of these points of discussion is that rotational instabilities may be excited in the whole of the fluid core by the differentially precessing mantle and inner core \cite<e.g.>{tilgner15,lebars15}.  Viscous dissipation may then occur in the volume of the fluid core in addition to that from friction at its solid boundaries. If so, this would contribute a part to the observed $\phi_p$, and our estimates of $Q_{cmb}$ and $Q_{icb}$ would be reduced.  Likewise, viscous relaxation can also occur within the solid inner core.  As shown in OD20, this requires that significant relaxation takes place over a timescale of one month, and in turn, this requires inner core viscosities in the range of $10^{13}-10^{15}$ Pa s.  If this is the case, the deformed inner core shape keeps a greater alignment with the mantle, reducing the amplitude of the periodic gravity signal induced by the inner core.  A lunar model with a large but viscously deforming inner core may then not generate a $|\Delta C_{21}|$ in excess of $5 \times 10^{-11}$ and would not be excluded. However, this would also imply that viscous relaxation within the inner core is taking a substantial share of the non-tidal part of the total dissipation and this also implies that $Q_{cmb}$ and $Q_{icb}$ would be reduced.

The estimate of the present-day dissipation in the core provides an anchor point to predict how it may have changed when projected back in time.  For instance, $Q_{cmb}$ was larger in the past because the rotation rate was faster and because the offset between the spin vectors of the mantle and fluid core was also larger.  Previous reconstructions show that $Q_{cmb}$ was possibly sufficiently large in the past to power a lunar dynamo \cite{williams01,dwyer11,cebron19}.  These reconstructions assume that the non-tidal part of the dissipation in the present-day Moon is solely due to viscous friction at the CMB. If an important fraction of the viscous dissipation occurs at the ICB, the total viscous dissipation projected back in time could be either enhanced or reduced (compared to model with $Q_{icb}=0$) depending on how $\Omega_{ficn}$ has evolved relative to $\Omega_p$.  An episode of enhanced dissipation would result if a crossing of the FICN resonance took place.  Viscous dissipation would be reduced prior to the nucleation of the inner core.  Such changes in past viscous dissipation would affect reconstructions of the power available to drive a lunar dynamo through time.  Furthermore, they would also influence the evolution of the lunar orbit inclination and therefore have an impact on reconstructions of the evolution of the Earth-Moon system \cite<e.g.>{williams01,cuk16,cuk19}.

The maximum present-day dissipation at the ICB that we find, approximately $7.7 \times 10^7$ W, is much weaker than the values computed in \citeA{stys20}.  Their results suggest that $Q_{icb}$ in the present-day Moon could be in excess of $10^{11}$ W in the vicinity of the FICN resonance and for a large inner core (see for instance their Figure 6e).  The manner by which $Q_{icb}$ is built in \citeA{stys20} is by, first, inferring a friction parameter at the CMB $f_{cmb}$ (which is equal to $0.75 \pi^2 \kappa_{cmb}$ in our notation) compatible with LLR observations, and second by assuming the same numerical value for $f_{icb}$.  This yields $f_{cmb}=f_{icb}$ in the range of  0.002-0.005 depending on the CMB radius (corresponding to $\kappa_{cmb}$ in the range of 0.00027 to 0.00068). However, as we demonstrate in Figure \ref{fig:sol}f, for a large inner core such a procedure leads to a phase lead which is largely in excess of the observed $\phi_p=0.27$ arcsec.  In other words, the method used in \citeA{stys20} is inconsistent with the present-day rotational dissipation observed through $\phi_p$.  As we show here, when the dissipation is constrained to match the observed $\phi_p$, $\kappa_{cmb}$ must be reduced with increasing inner core size, the more so when $\Omega_{ficn}$ is close to $\Omega_p$.  This adjustment, in turn, leads to a much smaller dissipation at the ICB.  In short, the total dissipation in the core is constrained by $\phi_p$, so $Q_{icb}$ cannot exceed the numerical value of $Q_{cmb}$ inferred in the absence of an inner core, which is approximately equal to $8.18 \times 10^7$ W.     The significantly reduced upper bound in the present-day $Q_{icb}$, by 3 to 4 orders of magnitude, implies that its projection back in time must be decreased by a similar factor.  Consequently, in contrast to the suggestion made in \citeA{stys20}, viscous dissipation at the ICB is unlikely to have ever been above the threshold to power a thermally driven lunar dynamo.  This, however, does not exclude the possibility that the mechanical stirring of core flows by a differentially precessing inner core may be capable of generating dynamo action.

A limitation of our rotational model is that it is built under the assumption of small angles of tilt.  It becomes largely inaccurate when the the FICN frequency approaches the orbital precession frequency and the inner core tilt is resonantly amplified to large angles.  An improvement would be to include tidal and viscous dissipation in a rotational model valid for all angles of tilt, for instance similar to that presented in SD18.  In such a model, the maximum inner core tilt would remain bound when $\Omega_{ficn}\approx\Omega_p$ and predictions of viscous dissipation at the ICB close to the FICN resonance would be improved.  


\appendix 
\section{Prediction of the phase lead angle}

We show here how the prediction for the phase lead angle $\phi_p$ given by Equation (\ref{eq:varphip_pred1}) is constructed.  It is based on the angular momentum equation for the whole Moon, given by the first row of Equation (A3) of OD20.  By using $\omega = -1 -\delta \omega$ and  $\tilde{m} = \delta \omega \tilde{p}$ (row 5 of Equation A3 of OD20), we can write this equation as

\begin{equation}
(\delta \omega M_{11} + M_{15}) \tilde{p} + M_{12} \tilde{m}_f + M_{13} \tilde{m}_s + M_{14} \tilde{n}_s = y_1 \, . \label{eq:app1}
\end{equation}
The mathematical expressions for the different matrix elements $M_{1j}$ and the right-hand side $y_1$ are given in Appendix A of OD20.  

We can simplify the expressions for each of the $M_{1j}$ and of $y_1$, first by setting all  compliances $S_{ij}$ equal to zero except for $S_{11}$, as appropriate for the rotational model that we use in the present study.  This gives

\begin{subequations}
\begin{align}
M_{11}  & = - 1 - \delta \omega - e + S_{11}(-\delta \omega + \Phi_2) \, \\  
M_{12} & = -  \frac{\bar{A}_f}{\bar{A}} \delta \omega \, ,\\
M_{13} & = -  \frac{\bar{A}_s}{\bar{A}} \delta \omega \, ,\\
M_{14} & =  \frac{\bar{A}_s}{\bar{A}} \alpha_s \big(-\delta \omega e_s + \beta_s \Phi_2 \big)  \, ,\\
M_{15} & = \Phi_2 \Big( \beta  +\delta \omega S_{11}  - 3 {\cal M} Re[S_{11}] \Big) - i Im[S_{11}] \Phi_2^t \, , \\
y_1 & =- \Phi_1 \Big( \beta  + \delta \omega  S_{11} - 3 {\cal M} Re[S_{11}]\Big) + i Im[S_{11}] \Phi_1^t  \, .
\end{align}
\label{eq:defs}
\end{subequations}
The definition of each variable is given in OD20.  Note that the Poincar\'e number ($\delta \omega = 4.022 \times 10^{-3}$), and the parameters involved in the gravitational torque from Earth (${\cal M}=0.9878$, $\Phi_1=0.1329$, $\Phi_2=1.4646$, $\Phi_1^t=0.4190$ and $\Phi_2^t=4.6895$) are independent of the interior lunar model.  The dynamical ellipticities of the whole Moon ($e$, $\beta$) and of the inner core  ($e_s$, $\beta_s$) depend on the choice of interior model, and are of the order of $10^{-4}$.  The factor $\alpha_3 = 1 - \rho_f/\rho_s$ also varies with the interior model through the densities of the fluid ($\rho_f$) and solid ($\rho_s$) cores, and is between 0 and 1.  The numerical value of the complex compliance $S_{11}$ is set by the choices of $k_2$ and $Q$ (see Equation \ref{eq:k2Q}).  With $k_2=0.02422$ and $k_2/Q= 6.4 \times 10^{-4}$, this gives $Re[S_{11}]=1.55 \times 10^{-7}$ and $Im[S_{11}]= 4.11 \times 10^{-9}$.  

Neglecting small terms, but keeping the largest terms involving $Im[S_{11}]$, and inserting each of the expressions in Equation (\ref{eq:defs}) in Equation (\ref{eq:app1}), we obtain

\begin{align} 
\Big( - \delta \omega (1 + \delta \omega + e) + \beta \Phi_2  - i Im[S_{11}] \Phi_2^t \Big) \tilde{p} & = \nonumber\\ 
& \hspace*{-4cm} \delta \omega\left(  \frac{\bar{A}_f}{\bar{A}}\tilde{m}_f + \frac{\bar{A}_s}{\bar{A}} \tilde{m}_s \right)   - \frac{\bar{A}_s}{\bar{A}} \alpha_3 \beta_s \Phi_2 \tilde{n}_s  -\beta \Phi_1 - i Im(S_{11}) \Phi_1^t \, . \label{eq:app2}
\end{align}
Taking the imaginary part of this equation, with $\phi_p = -Im[\tilde{p}]$, substituting $Im[S_{11}]$ with its expression in terms of $k_2/Q$ given by Equation (\ref{eq:k2Q}), and writing $\Phi^t =  \Phi_1^t + \Phi_2^t Re[\tilde{p}] \approx 0.5453$, one obtains a prediction for $\phi_p$ given by

\begin{equation}
\phi_p =  \left( \frac{1}{\delta \omega (1+e+ \delta \omega) - \beta \Phi_2} \right) \Bigg[ \left( \frac{k_2}{Q} \right) \frac{R^5\,  \Omega_o^2 \, \Phi^t }{3 G \bar{A}}  + \delta \omega\left(  \frac{\bar{A}_f}{\bar{A}} Im[\tilde{m}_f] + \frac{\bar{A}_s}{\bar{A}} Im[\tilde{m}_s] \right) - \frac{\bar{A}_s}{\bar{A}} \alpha_3 \beta_s \Phi_2 Im[\tilde{n}_s] \Bigg] \, .\label{eq:app3}
\end{equation}  
The value of $\phi_p$ from this approximate expression differs by less than 1 part in $10^3$ from that computed by our rotational model.

%
%
%
%
%
%
%
%

\acknowledgments
Comments and suggestions by two reviewers significantly improved this paper.  All Figures were created with the GMT software \cite{gmt}.  MD is supported by a Discovery Grant from NSERC/CRSNG.  The source codes, GMT scripts and data files to reproduce all figures are freely accessible at the following Dataverse repository:\\ https://doi.org/10.7939/DVN/AYCIRT.


%
%


%
%
%
%
%

\end{document}